\documentclass[%
aip,
amsmath,amssymb,
reprint,%
]{revtex4-1}

\usepackage{graphicx}
\usepackage{dcolumn}
\usepackage{bm}

\usepackage[utf8]{inputenc}
\usepackage[T1]{fontenc}
\usepackage{mathptmx}
\usepackage{braket}

\begin{document}
	
	\preprint{AIP/123-QED}
	
	\title{Polarization interferometric prism: a versatile tool for generation of vector fields, measurement of topological charges and implementation of a spin-orbit Controlled-Not gate}
	
	\author{Zhi-Cheng Ren}
	\author{Zi-Mo Cheng}
	\author{Xi-Lin Wang}
	\email{Author to whom correspondence should be addressed: xilinwang@nju.edu.cn}
	\author{Jianping Ding}
	\author{Hui-Tian Wang}
	\email{htwang@nju.edu.cn}
	\affiliation{National Laboratory of Solid State Microstructures and School of Physics, Nanjing University, Nanjing 210093, China}
	\affiliation{Collaborative Innovation Center of Advanced Microstructures, Nanjing 210093, China}
	
	\date{\today}

\begin{abstract}
Optical vortex and vector field are two important types of structured optical fields. Due to their wide applications and unique features in many scientific realms, the generation, manipulation and measurement of such fields have attracted significant interest and become very important topics. However, most ways to generate vector fields have a trade-off among flexibility, efficiency, stability, and simplicity. Meanwhile, an easy and direct way to measure the topological charges, especially for high order optical vortex, is still a challenge. Here we design and manufacture a prism: polarization interferometric prism (PIP) as a single-element interferometer, which can conveniently convert an optical vortex to vector fields with high efficiency and be utilized to precisely measure the topological charge (both absolute value and sign) of an arbitrary optical vortex, even with a high order. Experimentally we generate a variety of vector fields with global fidelity ranging from 0.963 to 0.993 and measure the topological charge of an optical vortex by counting the number of petals uniformly distributed over a ring on the output intensity patterns. As a versatile tool to generate, manipulate and detect the spin-orbital state of single photons, PIP can also work in single-photon regime for quantum information processing. In experiment, the PIP is utilized as a spin-orbit Controlled-Not gate on the generated 28 two-qubit states, achieving the state fidelities ranging from 0.966 to 0.995 and demonstrating the feasibility of the PIP for single photons. 
\end{abstract}

\maketitle

Optical vortex has a helical phase of $\exp (j m \varphi)$ and carries an orbital angular momentum (OAM) of $m\hbar$ per photon,\cite{Allen1992} where $\varphi$ is the azimuthal angle and the integer index $m$ is the topological charge. As one of the most important structured light,\cite{Forbes2019, Yuan2019} optical vortices keep attracting great attention, and their generation and detection develop fast ranging from optical~\cite{Forbes2016, Wang2007} to acoustic~\cite{Zhang2019} and radio frequency.\cite{Mao2018} Till now, optical vortices have already contributed in quantum information,\cite{Alois2001} optical tweezers,\cite{He1995} and microscopy,\cite{Willig2006} etc. Meanwhile, optical vortices promote the emergence and development of optical vector fields with spatially-variant polarization state~\cite{Zhan2009, Milione2011}, which are another important type of structured light. The special polarization distributions of vector fields result in some peculiar properties in the research of photons themselves and light-matter interaction, for example, shaper focus,\cite{Dorn2003} longitudinally polarized optical needles,\cite{Wang2008} polarization Möbius strips,\cite{Bauer2015} and so on. Vector fields have found a wide variety of applications in microscopy,\cite{Kozawa2018} nonlinear optics,\cite{Camacho2016} material processing~\cite{Weber2011} and even quantum information.\cite{Barreiro2008} 

Many methods have been developed to generate vector fields, such as interferometric scheme,\cite{Wang2007, Forbes2016} intra-cavity,\cite{Naidoo2016, JLZhao2017} $q$-plate,\cite{Marrucci2006} meta-vector-polarizer,~\cite{Bomzon2002} and nanostructure.\cite{Beresna2011} Laser cavity, $q$-plate, meta-vector-polarizer and nanostructure are lack of flexibility. With the help of spatial light modulator (SLM), interferometric schemes are quite flexible, but are relatively complicated. Considering the close connection between optical vortex and vector field, a simple and stable device, which can conveniently and efficiently convert an optical vortex into vector fields, will be of great value. On the other hand, most common ways to measure the topological charge of optical vortex is to use its diffraction or/and interference properties, such as slit diffraction and interference,\cite{Ferreira2011, Zhou2014} interference with plane waves,\cite{Huang2013} Mach--Zehnder or Michelson interferometer,\cite{Li2016,Pushin2020} diffraction by an aperture\cite{Hickmann2010,CSGuo2009} and grating,\cite{JLZhao2018} and cylindrical lens.\cite{LXChen2016, Alperin2016, BGu2020} These methods provide more choices for measuring the OAM number. A compact and stable scheme will enrich the tools for the OAM research. In this letter, we design a single-element interferometer with a compact and stable scheme, which is quite convenient in generating a variety of vector fields, measuring the topological charge of optical vortex, and manipulating the spin-orbital state of single photons.

Vector fields could be expressed as the coherent superposition of two orthogonally polarized scalar fields with opposite phases. The key is to produce two opposite vortices in two paths, which can be usually achieved by optical fields in one path being reflected even times and the other odd times.\cite{Mendoza2019} To this end, we design and develop a single-element interferometer, named as polarization interferometric prism (PIP) in Fig.~1(a). The PIP is composed of two fused silica prisms (A and B) glued together in Fig.~1(b). The polarization splitting coating on the glued surface (60 mm in length) of the prisms A and B divides the input beam into two beams with orthogonal linear polarizations in the lower left corner and combine them in the upper right corner. As shown in Fig.~1(b), four angles in the lower left and upper right corners for the prisms A and B are all designed to be $45^\circ$, and the other three corners (A1, A2 and A3) for the prism A have the angles ($157.5^\circ$, $135^\circ$ and $157.5^\circ$) and the other four corners (B1, B2, B3 and B4) for the prism B have the angles ($165^\circ$, $150^\circ$, $150^\circ$ and $165^\circ$). The PIP has a thickness of 5 mm, which offers an aperture of $\sim$5 mm. The horizontally (vertically) polarized component is transmitted (reflected) on the glued surface of the PIP, and will undergo 2 (3) total internal reflections at the surface of the prism A (B) before combination. The two optical paths inside the PIP are carefully designed to be equal to ensure the perfectly coherent superposition. Obviously, the PIP is immune to air turbulence or/and table vibration, so the PIP is a stable single-element interferometer even for poor-coherence lasers.

\begin{figure}[!ht]
\centering
        \includegraphics[width=8.0cm]{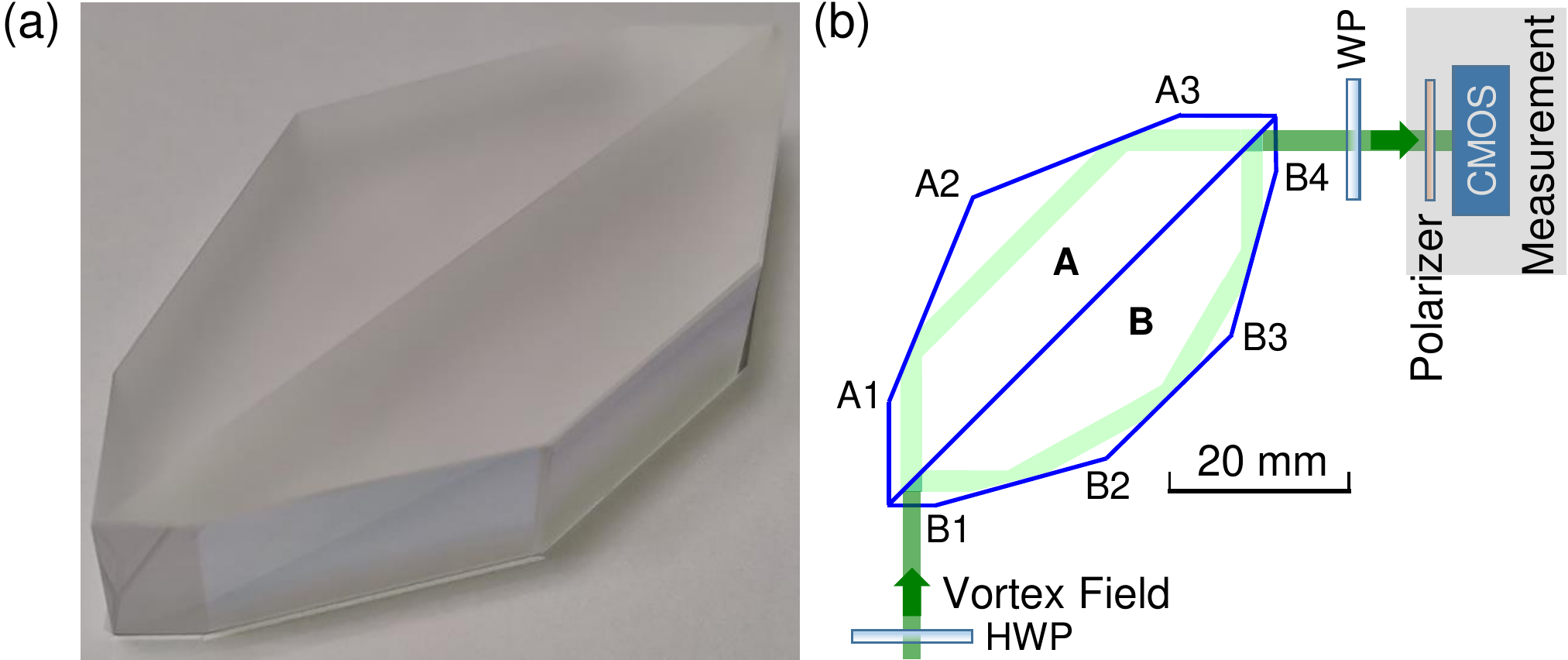}
\caption{Generator of vector fields and topological charge measurement of optical vortices, with a single-element interferometer (PIP). (a) Photo of a PIP. (b) Converter of optical vortex to vector field and measuring instrument of topological charges for optical vortices. HWP, half-wave plate is used to adjust intensity ratio in two paths inside the PIP. WP, wave plate is used to select the superposition bases for vector field.}
\end{figure}

To generate optical vortices, many methods have already proposed, such as using a spiral phase plate (SPP) or computer-generated holograms.\cite{Forbes2019} To verify the feasibility of our device, we use a commercial spatial light modulator (SLM) to experimentally generate optical vortex $\exp (j m \varphi )$ with arbitrary topological charge $m$. As mentioned above, the two beams propagating inside the PIP undergo even and odd total internal reflections respectively, so the two orthogonal linearly polarized beams output from the PIP will have opposite helical phases $\exp ( \pm j m \varphi )$. We use a Soleil-Babinet compensator composed of two wedge-shape wave plates to precisely control the relative phase delay between the two beams.

When the fast axis of the half-wave plate (HWP) in front of the PIP is oriented at $22.5^\circ$ (all orientation angles are defined with respect to the horizontal direction), the two beams in the two paths have the balanced intensity; when the wave plate (WP) behind PIP is a quarter-wave plate (QWP) and its fast axis is oriented at $45^\circ$, the two orthogonal linearly polarized beams output from the PIP are converted into the orthogonal circularly polarized beams. Under these conditions, the generated cylindrical vector optical field can be written as 
\begin{equation}
\textbf{E} \propto \exp (- j m \varphi ) \hat{\textbf{e}}_R + \exp  (j m \varphi ) \hat{\textbf{e}}_L,
\end{equation}
where $ \hat{\textbf{e}}_R$ ($\hat{\textbf{e}}_L$) refers to the right-handed (left-handed) circularly polarized unit vector. Such a kind of vector fields is locally linearly polarized and can be represented by a point on the equator of a higher-order Poincar\'{e} sphere.\cite{Milione2011} 

When the input optical vortex has the topological charge of $m = 1$, the generated radially polarized vector field is shown in the first column of Fig.~2(a). To make the polarization distributions more obvious and observable, we expand the input beam to homogeneous intensity in Fig.~2. We can easily adjust the phase delay between the two components to be $ \pi $ and generate the azimuthally polarized vector field. When the input vortex fields have higher-order topological charges of $m = 3$ and $5$, the generated high-order cylindrical vector fields are shown in the third and fourth columns of Fig.~2(a). We can use a horizontal polarizer to distinguish the vector fields. Behind a polarizer, the uniform intensities become into the fan-shaped patterns with $2|m|$ extinction lines where the local linear polarizations are orthogonal to the polarizing direction of the polarizer in the second row of Fig.~2(a). 

Our device can also generate hybrid vector fields and arbitrary higher-order Poincaré sphere states,\cite{Milione2011} which can be generated by the setup in Fig.~1. To fully characterize the polarization state distributions, Stokes parameters of $S_1$, $S_2$ and $S_3$ are also measured. As shown in the first row of Fig.~2(b), the hybrid vector field with azimuthal-dependent Stokes parameters of $S_1$, $S_2$ and $S_3$ is created by orthogonal elliptically polarized bases of $\{ \sin\alpha \hat{\textbf{e}}_H \! + \! \cos\alpha e^{j\beta} \hat{\textbf{e}}_V, \cos\alpha \hat{\textbf{e}}_H \! - \! \sin\alpha e^{j\beta} \hat{\textbf{e}}_V \}$ with $\alpha \! = \! 52.2^\circ$ and $\beta = - 0.35 \pi$, where $\hat{\textbf{e}}_H$ ($\hat{\textbf{e}}_V$) is the horizontally (vertically) unit vector. This hybrid vector field can be generated when the WP behind PIP is a rotatable HWP.

The higher-order Poincaré sphere vector fields can also be generated as
\begin{equation}
\textbf{E} \propto a \exp (-j m \varphi) \hat{\textbf{e}}_R + b \exp (+j m \varphi) \hat{\textbf{e}}_L, 
\end{equation}
where $a$ and $b$ indicate the relative intensities with $a^2 + b^2 \equiv 1$, and are easily adjusted by rotating the HWP in front of PIP. When the HWP is oriented at $43.56^\circ$ (i.e. $a = 0.358$ and $b = 0.934$), the horizontally and vertically polarized components have an intensity ratio of $13 : 87$ for the input beam. As shown in the second row of Fig.~2(b), the generated higher-order Poincaré sphere vector field with $m = 1$ has homogeneous nonzero $S_3$ but azimuthal-dependent $S_1$ and $S_2$. 
  
 \begin{figure}[!ht]
 \centering
         \includegraphics[width=8.0cm]{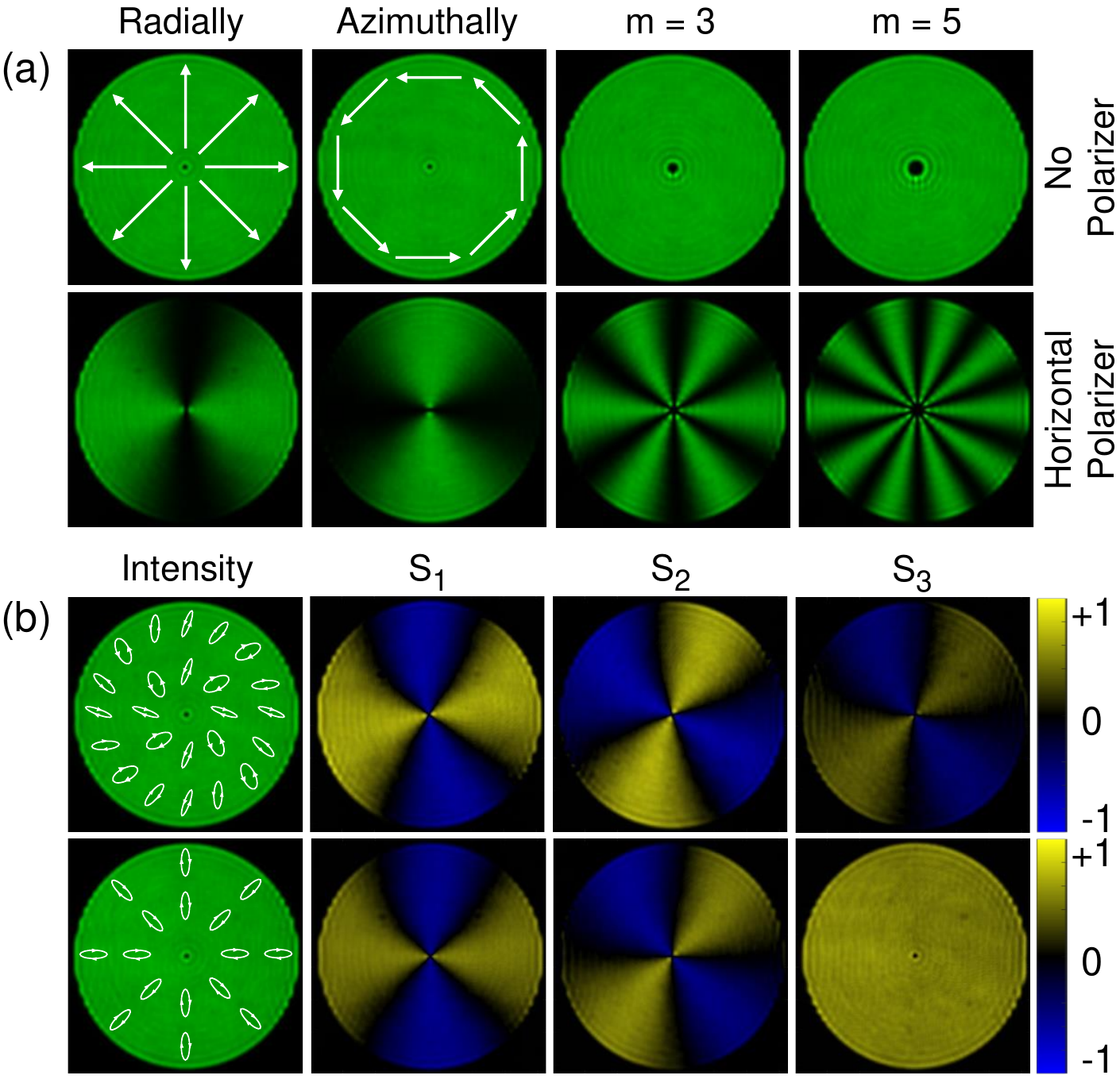}
 \caption{(a) Generated four vector fields with local linear polarization; (b) generated hybrid and higher-order Poincaré sphere vector fields, and corresponding Stokes parameters.}
 \end{figure}
 
Local Stokes parameters in Fig.~2(b) are difficult to directly characterize the global quality of the generated vector fields. It will be useful to evaluate the global quality of the generated vector field by a single parameter. Here we refer to the concept of fidelity,\cite{Nielsen2010} which can characterize the global similarity or difference between the experimentally generated state and the target state. The fidelity is defined as $ F(\rho_t, \rho_e) \equiv \mathrm{Tr} (\rho_t, \rho_e)$, where $\rho_t = |E \rangle \langle E| $ ($\rho_e = |E_e \rangle \langle E_e| $) represents the density matrix of the target vector field state (the experimentally generated vector field state). We can rewrite the density matrix with the Stokes parameters as follows 
\begin{equation}   
\rho =    
\frac{1}{2} \left[               
  \begin{array}{cc}   
   1 + S_1 & S_2 + j S_3 \\  
   S_2 - j S_3 & 1-S_1 \\  
  \end{array}
\right].                 
\end{equation}
With the Stokes parameters normalized by the total intensity $S_0$ (satisfying $S_1^2+S_2^2+S_3^2=1$), the fidelity of the complete polarized vector field we generated can be rewritten as 
\begin{equation}   
   F(\rho_t, \rho_e) = (1+S_1^t S_1^e+S_2^t S_2^e+S_3^t S_3^e)/2,               
\end{equation}
where $S_i^t$ and $S_i^e$ ($i = 1,2,3$) refer to the normalized Stokes parameters of the target state and the experimentally generated vector field state, respectively. By averaging the local fidelities $F(\rho_t, \rho_e)$, we can obtain the global fidelity as $F_{global}(\rho_t, \rho_e)$ to characterize the quality of the generated vector field. A large number of vector fields we generated have very high quality, because they have the global fidelities ranging from 0.963 to 0.993.

Obviously, it can be seen from Fig.~2 that behind a polarizer, the uniform intensity turns into fan-shaped patterns with $2|m|$ bright petals. This offers an intuitive way to measure the topological charge of an optical vortex. By using the same setup in Fig.~1, as examples, Fig.~3 shows the measured optical vortices with topological charges of $m=+4$, $-16$, $+64$ and $+65$. One should point out that in order to eliminate the unwanted subsidiary rings in the focal plane, as shown in Fig.~3, we use the optimal annulus structure mask for generating the input vortex,\cite{Guo2004} which shows the fact that the optimal annulus width of the input vortex depends on the topological charge. As shown in the first two columns of Fig.~3, the absolute values of topological charges are easily determined by counting the bright petals at either the image plane or the focal plane. In details, the patterns in the first two columns show 8, 32, 128 and 130 bright petals, respectively. Correspondingly, we can identify the absolute values of topological charges to be $| m | = 4$, 16, 64 and 65, respectively. However, the sign of topological charge cannot be directly obtained, due to the same number of petals ($2|m|$) for the optical vortices with $ \pm m$. To solve this problem, an auxiliary measurement should be carried out: inserting a spiral phase plate (SPP) with a phase structure of $\exp (j m' \varphi)$ into the input port of the original setup. As an example with $m' \! \! = \! +1$, the second measured patterns are shown in the last two columns of Fig.~3. Comparing the two measured patterns, the sign of topological charge can be easily determined. If the number of petals increases by 2, the topological charge is positive. Otherwise, if it decreases by 2, the topological charge is negative. By counting the number of petals, it is easy to distinguish $m=+64$ from $m=+65$, which proves the high accuracy and convenience of our method.

\begin{figure}[!ht]
\centering
        \includegraphics[width=8.0cm]{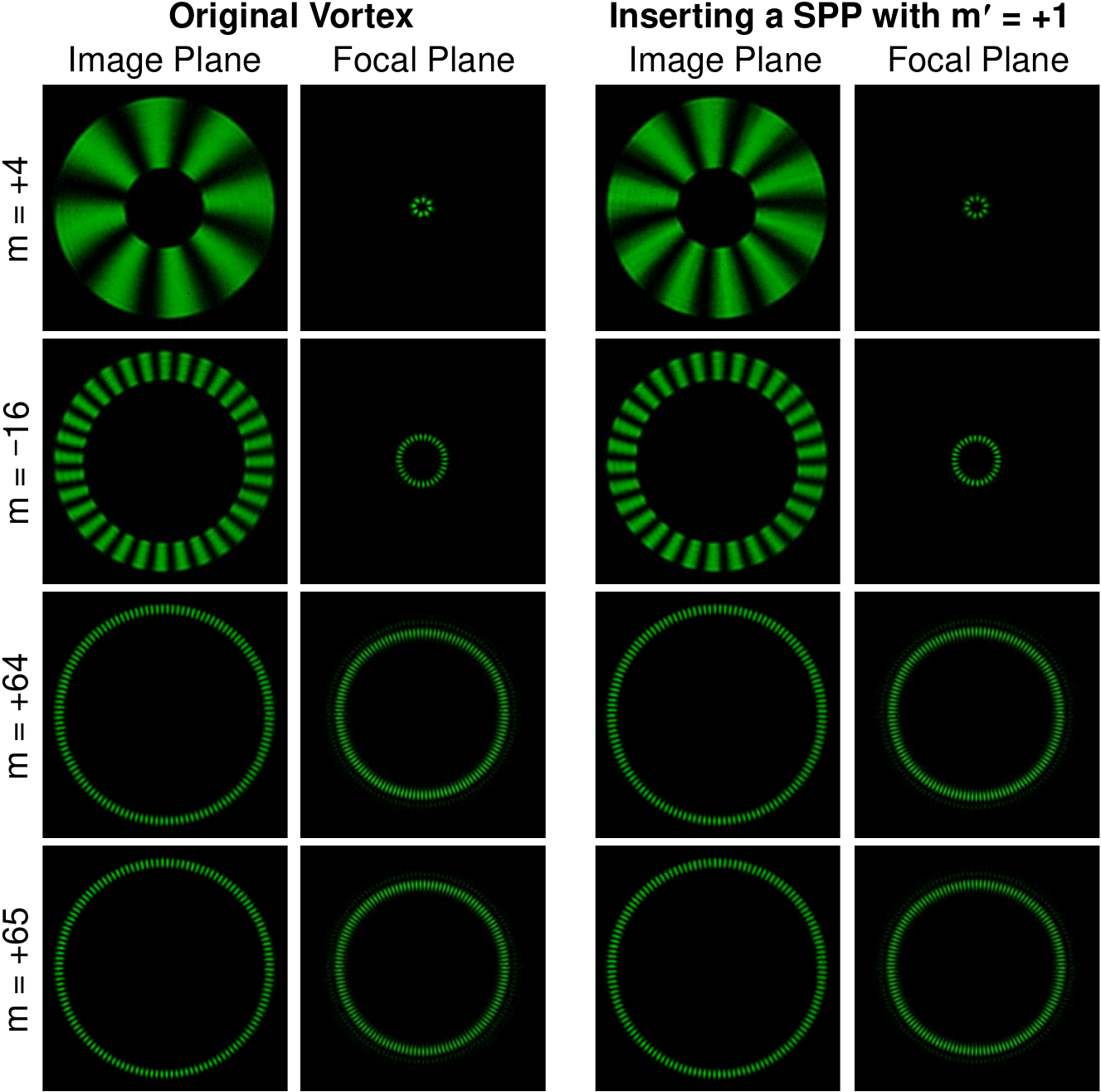}
\caption{Measurement of topological charges of optical vortices with $m=+4$, $-16$, $+64$ and $+65$. The first and second columns show results using the same scheme in Fig.~1, while the third and fourth columns show the results by inserting a SPP with $m' \! \! = \! +1$ into the input port of the original setup.}
\end{figure}

The above two applications of PIP, namely generation of optical vector fields and measurement of topological charges, require to record the whole image of optical fields by exposing a large number of photons. A question to be asked is whether this PIP can also work in single-photon regime. Our PIP can implement a spin-orbit Controlled-NOT (CNOT) gate for single photons, with the spin or polarization as the control qubit and the OAM as the target qubit. To verify this functionality, we need to prepare a single-photon 4-dimension spin-orbit composite state $a {\ket H} \ket {+1} + b {\ket V} \ket {+1} + c {\ket H} \ket {-1} + d {\ket V} \ket {-1}$, where ${\ket H}$ (${\ket V}$) denotes horizontal (vertical) polarization and $\ket {+1}$ ($\ket {-1}$) denotes positive (negative) OAM of $+ \hbar$ ($- \hbar$). As shown in Fig.~4(a),  we generate a heralded single photon, which is one of a photon pair from spontaneous parametric down conversion (SPDC) by passing a femtosecond (fs) pulsed laser through a 0.6 mm type-I $\beta$-barium borate (BBO), and triggered by detecting its twin photon. The heralded single photon is filtered with a single-mode fiber to obtain fundamental Gaussian mode (zero OAM mode). A $q$-plate and several WPs are utilized to prepare a spin-orbit composite state. Then this photon passes through PIP and a spin-orbit CNOT gate operates on its state. The converted state of two qubits is carried by a single photon and we measure it in a step-by-step manner, first polarization measurement then OAM measurement\cite{WangXL2018}. One of the most important functions of a CNOT gate is to transfer two qubit product states to entangled states and vice versa. Here we convert the 4 Bell states of $\ket{\Phi^{\pm}}=({\ket H} \ket {+1} \pm{\ket V} \ket {-1})/\sqrt{2}$ and $\ket{\Psi^{\pm}}=({\ket H} \ket {-1} \pm{\ket V} \ket {+1})/\sqrt{2}$ into the 4 product states with very high fidelity and vice versa in Fig.~4(b). Moreover, we also implement the spin-orbit CNOT gate on all 20 states in 5 groups of mutually unbiased basis~\cite{Wiesniak2018} and achieve high fidelity for the converted state ranging from 0.966 to 0.995, demonstrating the feasibility of our PIP for single photons.

\begin{figure}[!ht]
    \centering
        \includegraphics[width=8.0cm]{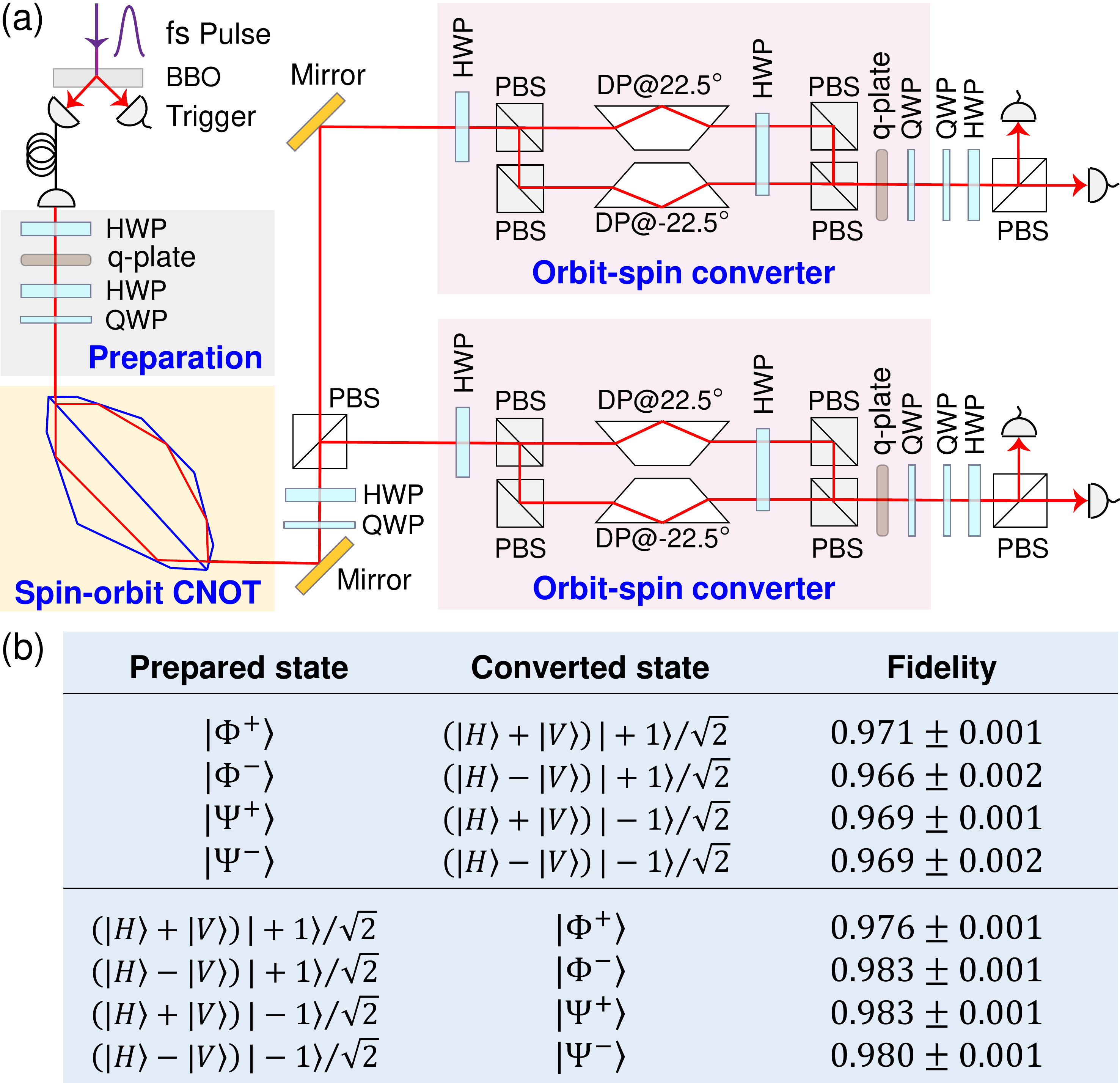}
\caption{Experimental setup and results for confirming the feasibility of PIP for single photons. (a) PIP, as a CNOT gate, operates on a 4-dimension spin-orbit composite state prepared by a heralded single photon, and the converted state is firstly measured in polarization basis, then in OAM basis. (b) The measured fidelities for the converted 4 Bell states and the converted states from the prepared 4 Bell states. }
\end{figure}   

In summary, we have designed and developed a PIP, which can convert vortex fields into vector fields with simultaneously high quality, low complexity and high efficiency. Our single-element PIP enables vector fields more widely used, especially at high power and high energy. Based on PIP, we also propose a method to conveniently and precisely measure the topological charge of any optical vortex. Compared with previous methods, our PIP has the advantages of compact, high stability, high efficiency, and great convenience. More importantly, the number of uniformly distributed petals is easier to observe and count than diffracted stripes or nonuniform spots, especially for large topological charges. Benefiting from its high stability, the PIP has the advantage for large-scale applications with dozens of single-photon interferometers.\cite{WangXL2018} Furthermore, our PIP can also work at single-photon level and can be utilized as a versatile quantum element to generate, manipulate and detect the spin-orbital state of single photons. 

\

This work was supported by the National Key R\&D Program of China (2017YFA0303800, 2017YFA0303700, 2018YFA0306200); National Natural Science Foundation of China (11534006, 91750114, 11922406, 91750202).

\

\noindent \textbf{DATA AVAILABILITY}

\ 
The data that support the findings of this study are available from the corresponding author upon reasonable request.

\end{document}